\documentclass[pra,aps,twocolumn,eqsecnum,showpacs]{revtex4}
\usepackage{mathtext}
\usepackage[T2A]{fontenc}
\usepackage[cp1251]{inputenc}
\usepackage{epsf}
\usepackage{psfrag}
\usepackage{graphicx}
\usepackage{amssymb,amsmath}

\begin{document}

\title{Spatial distribution of optically induced atomic  excitation in a dense and cold atomic ensemble}
\author{Ya. A. Fofanov${}^{1}$, A. S. Kuraptsev${}^{2}$, and I. M. Sokolov${}^{1,2}$\\
{\small $^{1}$Institute for Analytical Instrumentation, Russian Academy of
Sciences, 190103, St.-Petersburg, Russia }\\
{\small $^{2}$Department of Theoretical Physics, State Polytechnic
University, 195251, St.-Petersburg, Russia }\\
M. D. Havey\\
{\small Department of Physics, Old Dominion University, Norfolk, VA 23529}}


\sloppy



\begin{abstract}
On the basis of our general theoretical results developed previously in JETP 112, 246 (2011), we calculate the spatial distribution of atoms excited in a dense and cold atomic cloud by weak monochromatic light. We also study the atomic distribution over different Zeeman sublevels of the excited state in different parts of the cloud.  The dependence of this distribution of atomic excitation on the density of the atomic ensemble and the frequency of external emission is investigated. We show that in the boundary regions of the cloud the orientation and alignment of atomic angular momentum takes place. Analysis of the spatial distribution of atomic excitation shows no noticeable signs of light localization effects even in those parameter regimes where the Ioffe-Regel criterium of strong localization is satisfied.  However, comparative calculations performed in the framework of the scalar approximation to the dipole-dipole interaction reveals explicit manifestation of strong localization under some conditions.

\end{abstract}


\maketitle

\section{Introduction}
Cold and ultra-cold atomic ensembles are currently of significant interest for a wide range of practical applications in metrology, development of frequency standards, the possibility of cold-atom based random lasing
\cite{Cao,Wiersma,Conti,KaiserRandom1,KaiserRandom2}, and quantum information science \cite{3,4,5,6,7,8,9,10,11}. The efficacy of such practical applications depends substantially on the optical thickness of these ensembles. In the overwhelming majority of the discussed applications this thickness must be large. An increase in the thickness can be realized either by means of an increase in the dimensions of the ensemble or with an increase in its density. Improvements in techniques for cooling atomic gases in atomic traps has permitted development of atomic samples of substantial density. Such dense atomic clouds, in which the average interatomic distances are comparable with the optical wavelength, have attracted much attention lately. We have seen that, besides practical applications mentioned above, interest in dense ensembles is generated by their unique, and sometimes surprising, physical properties.

One of the most interesting effects which is expected to be observed in cold dense atomic ensembles is Anderson (strong) localization of light \cite{24,25,26,27}. This phenomenon consists of suppression of light propagation in a randomly heterogeneous medium and is caused by destructive interference of multiply scattered waves. According to one of the criteria, as formulated by Ioffe and Regel, strong localization is possible under the condition $ kl  < 1$, where $k$ is the wave number of the light, and $l$ is the mean free path of a photon in the medium. Thus, to achieve localization it is necessary to make the mean free path of a photon substantially less than the wavelength of light. Cold atoms are characterized by large resonance light scattering cross sections, which allows for small mean free paths to be obtained at high atomic density. Besides this, in cold ensembles the frequency of resonance photons remains near resonance even after multiple scattering. The Doppler frequency shift does not play such as important role as in hot gases, and recoil shifts are small even after a substantial order of scattering. For these reasons, and others, cold atomic ensembles have been considered as promising systems for experimental realization of light localization.

In spite of the successes of the creation of very dense and cold ensembles, their experimental study, and theoretical interpretation of measurements, remains challenging \cite{Mark_exp,Mark_exp2}. For this reason, the theoretical analysis of the dense ensembles remains an essential tool for studying the possibility of the existence of Anderson Localization in an atomic gas of point scatterers, as as well as optimum conditions of observation of this effect.

In the model of independent scatterers the theoretical determination of the conditions, which meet the Ioffe-Regel criterion for the cold atomic ensembles, was carried out in \cite{Kaiser}. This model does not take into account interatomic interactions and gives only approximate estimations of the parametric domain where strong localization may take place.  The analysis of the Ioffe-Regel condition based on a model of depending scattering was recently made in \cite{FKSH11}.

In this paper the atomic polarization created by weak monochromatic light in an optically thick, dense and cold atomic ensemble is calculated. It is shown that the amplitude of the polarization averaged over a uniform random atomic distribution decreases exponentially beyond the boundary regions. The phase of this polarization increases linearly in this region. On these grounds, the wavelength of the light in the dense atomic medium, its extinction coefficient, and the complex refractive index and dielectric constant of the medium were determined. Knowing these characteristics of the cold atomic clouds clouds allows us to fix the conditions within which strong localization, taking into account resonant dipole-dipole interatomic interaction, might take place.

Fulfilling of the Ioffe-Regel criterion is likely a necessary, but not sufficient condition for the observation of Anderson Localization. For instance, it does not guarantee the existence of strong localization.  A clearer answer to the question about light localization in cold gases can be obtained by means of analysis of the incoherent radiation transfer in such gases. For this purpose, in the present paper we calculate the spatial distribution of atoms excited in a dense cold atomic cloud by weak monochromatic light. Here the atomic ensemble is assumed to be homogeneous on average. Since in a homogeneous medium the excitation distribution is directly connected with the spatial distribution of the exciting radiation, this calculations allows us to draw qualitative conclusions about the influence of incoherent scattering and interference on radiation transfer in the ensemble.

Our calculations are based on the microscopic approach developed in \cite{SKH11}. Taking into account the vector properties of the scattered light and the internal structure of the atomic levels we calculate not only the total population of the excited atomic multiplet, but also the population of each Zeeman sublevel of this state.  We show that in the boundary regions of the sample, different sublevels are populated differently. In these regions orientation and alignment  of the atomic angular momentum take place. Polarization analysis gives us the opportunity to study depolarization of the trapped radiation.

Finally, we analyze the dependence of the spatial distribution of atomic excitation on the density of the atomic ensemble and the frequency of external excitation. We reiterate that in these calculations the vector model of the dipole-dipole interaction is fully implemented.  This analysis shows no noticeable signs of light localization effects even in those parameter regions where the Ioffe-Regel criterium of strong localization is satisfied.  However, a comparative calculation performed in the framework of the often-used scalar approximation to the dipole-dipole interaction displays explicit manifestation of strong localization for some conditions.

In the following paragraphs, we first lay out our basic assumptions and approach.  This is followed by presentation of our most important results, and a discussion of them.

\section{Basic assumptions and approach}
In the present work we analyze the optical steady state of an atomic ensemble consisting of $N$ atoms interacting with a quasi resonant monochromatic electromagnetic field. In our calculations we make the following assumptions. All atoms are identical and have a ground state $J=0$ separated by the frequency $\omega_a$ from an excited $J=1$ state. The natural linewidths of the three Zeeman sublevels of this state ($m=-1,0,1$) are $\gamma$.  Such level structure corresponds to Group II atoms, for instance, and makes it possible to consider correctly the vector nature of electromagnetic radiation, its polarization, and also selection rules for atomic transitions. Atoms are assumed to be motionless. The influence of the residual motion typical for cold clouds, prepared in the atomic traps, is taken into account by means of examining the statistical ensemble of clouds with the random distribution of atoms. The major part of the results presented below are obtained by averaging over this ensemble by the Monte Carlo method.

The microscopic approach enables us to consider the clouds of arbitrary form with an arbitrary nonuniform spatial distribution of atoms. However, taking into account the purpose of present work we consider only a cylinder-shaped cloud with a radius $R$ and the length $L$. The random distribution of atoms is assumed to be uniform on the average.

The external field is a plane monochromatic wave with frequency $\omega$
\begin{equation}
\mathbf{E}= \mathbf{e} E_0 exp(-i\omega t+i\mathbf{kr}),  \label{1}
\end{equation}%
Here  $\mathbf{e}$ and $E_0$ are the polarization vector and amplitude of the field. The polarization of the wave can be chosen arbitrarily. Further, for distinctness we will consider the incident field to be circularly polarized. The intensity of the radiation is assumed to be sufficiently small that all nonlinear effects are considered negligible. The wave vector $\mathbf{k}$ is directed along the axis of the cloud which will be denoted as the z-axis.

As is known, a weak monochromatic wave can be considered as a superposition of the vacuum and a small admixture of a one photon state. This approach, as well as the other approximations depicted above, allows us to avail ourselves of the results of the general theory developed in \cite{SKH11}. According to \cite{SKH11} the nature of atomic excitation is determined by the amplitudes of collective atomic states with one excited atom $b_{e_a^m}(t)$. These amplitudes can be found as follows

\begin{equation}
b_{e_a^m}(t)= b_{0}exp(-i\omega t)\sum\limits_{e_b^{m'}}
R_{e_a^m e_b^{m'}}(\omega))b^0_{e_b^{m'}}. \label{2}
\end{equation}%
Here, the index $e_a^m$ contains information about the number $a$ of atoms excited in any considered case. It also indicates the specific Zeeman sublevel $m$ which is populated.

The vector $b^0_{e_b^m}$ is determined by the initial excitation through the nonmodified external field
\begin{equation}
b^0_{e_b^m}= -\frac{\mathbf{d}_{e_b^m g_b}\mathbf{e}E_0}{\hbar},  \label{3}
\end{equation}%
where $\mathbf{d}_{e_b^m g_b}$ is the dipole matrix element for transitions from the ground $g$  to the excited $e^m$ state of atom b.

The matrix $R_{e_a^m e_b^{m'}}(\omega)$ in Eq. (\ref{2})  takes into account the possible excitation of the atoms by the trapped radiation. As was shown in \cite{SKKH09}, $R_{e_a^m e_b^{m'}}(\omega)$ is a resolvent operator of the considered system projected on the states consisting of single atom excitation, distributed over the ensemble, and the vacuum state for all the field modes
\begin{equation}
R_{e_a^m e_b^{m'}}(\omega )=\left[ (\omega -\omega _{a})\delta _{e_a^m e_b^{m'}}-\Sigma
_{e_a^m e_b^{m'}}(\omega )\right] ^{-1},  \label{4}
\end{equation}

The matrix $\Sigma_{e_a^m e_b^{m'}}(\omega )$ describes the excitation exchange between pairs of atoms inside the cloud. If $a\neq b$, its matrix elements are
\begin{eqnarray}
&&\Sigma _{e_a^m e_b^{m'}}(\omega )=\sum\limits_{\mu ,\nu}
\frac{\mathbf{d}_{e_a^{m};g_{a}}^{\mu }\mathbf{d}_{g_{b};e_b^{m'}}^{\nu }}{\hbar r^{3}}\times \label{5} \\&& \left[ \delta _{\mu \nu }\left(
1-i\frac{\omega _{a}r}{c}-\left( \frac{\omega _{a}r}{c}\right) ^{2}\right)
\exp \left( i\frac{\omega _{a}r}{c}\right) \right. -
\notag  \\&&
\left. -\dfrac{\mathbf{r}_{\mu }\mathbf{r}_{\nu }}{r^{2}}\left( 3-3i\frac{\omega _{a}r}{c}-\left( \frac{\omega _{a}r}{c}\right) ^{2}\right) \exp
\left( i\frac{\omega _{a}r}{c}\right) \right] .
\notag
\end{eqnarray}

Here $\mathbf{r}_\mu$ is the projection of the vector $\mathbf{r}=\mathbf{r}_{a}-\mathbf{r}_{b}$ on the axis $\mu$ of the chosen coordinate frame and $r=|\mathbf{r}|$ is the interatomic distance.

If $a$ and $b$ are the same atom then $\Sigma _{e_a^m e_b^{m'}}(\omega )$ differs from zero only for $m=m^\prime$. In this case $\Sigma _{e_a^m e_a^{m}}(\omega )$ determines the Lamb shift and the decay constant of the corresponding excited state. Including Lamb shifts in the transition frequency $\omega _{a}$ we get
\begin{equation} \Sigma _{e_a^m e_a^{m'}}(\omega )=-i\delta_{mm'}\gamma /2.
\label{6}
\end{equation}

Knowledge of explicit expressions for the amplitudes (\ref{2}) allows us to determine the spatial distribution of population of different Zeeman sublevels. Under the considered steady state conditions the population of sublevel $m$ in a unit volume can be calculated as follows
\begin{eqnarray}
\mathcal{\rho}_m(\mathbf{r})&=&\frac{1}{\Delta V}\left|\sum_{a\in \Delta V}
\sum\limits_{e_b^{m'}}R_{e_a^m e_b^{m'}}(\omega))b^0_{e_b^{m'}} \right |^2.\label{7}
\end{eqnarray}

In the next section, we use relation (\ref{7})  to calculate the spatial distribution of atomic excitation  and to analyze, on this foundation, incoherent light propagation through ensembles of different densities.

\section{Results and discussion}

\subsection{Vector model of the dipole-dipole interaction}

We begin our analysis with consideration of the total population of all excited Zeeman sublevels. The results of the corresponding calculations for a cloud with length $L = 10$ and radius $R = 15$ are shown in Fig. 1. Hereafter in this paper we use the inverse wavenumber of the resonant probe radiation in vacuum $k^{-1}=\lambdabar$. In these units, the mean density of atoms in this case is $n = 0.3$. For such a density, collective effects play an important role, particularly the Ioffe-Regel criterium is satisfied in some spectral interval (see \cite{FKSH11}).
\begin{figure}[th]
\begin{center}
{$\scalebox{0.85}{\includegraphics*{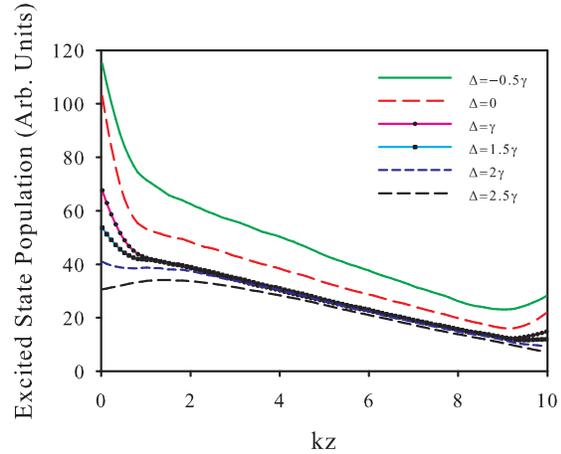}}$ }
\caption{Spatial distribution of atomic excitation. Calculations were made for a cylindrical cloud with length $L=10$ and radius $R=15$, atomic density is $n=0.3$. }
\end{center}
\par
\label{fig1}
\end{figure}

To avoid the influence of boundary effects at the lateral surface of the cylinder we calculate atomic excitation only for an area near the axis of the cylinder where we can neglect the dependence of the polarization on the transverse coordinate $r$. In this area we deal with a quasi one-dimensional case. The excitation depends only on z. Our analysis shows that for the considered parameters this take place for the inner portion of the cylinder with $r<10$. Results shown in Fig. 1 are obtained by averaging of the atomic excitation over the region with radius $r = 6$.

The main result demonstrated by Fig. 1 is the linear decrease of the excited state populations beyond the boundary regions. Note that this dependence differs significantly from the spatial distribution of the atomic polarization (see \cite{FKSH11}). The latter diminishes into the depths of the cloud exponentially. The basic reason for this difference lies in the fact that the total population of excited states is determined by the total intensity of light, and the polarization by the coherent component only.

At the same time the intensity of the radiation is not the only factor which determines the magnitude of the excited state populations. The lifetime of the collective multiatomic excited states plays a significant role.  Among these states there are both sub- and super-radiant states. Because of the dipole-dipole interaction different states are characterized by different resonance frequencies, and beyond that their spectral distribution is rather complicated. These circumstances explain the observed dependence of the value of atomic excitation on the frequency $\omega$ of the external radiation. The greatest total population of all excited states takes place for a detuning $\Delta=\omega-\omega_a$ which is about $-0.5 \gamma$. In this spectral region the density of the long-lived collective states is greatest.

In the central regions of the cylindrical volume  the medium is uniform on the average, and the spatial distribution of excited atoms is determined by the spatial distribution of the intensity of radiation. On these grounds it is possible to make some conclusions about the nature of the radiation transfer in the considered clouds. The linear decrease of the intensity of light, for instance, is characteristic for diffusive radiative transfer. There is a known relationship, which shows how the intensity of the light decreases insider the plane layer \cite{Ross}

\begin{equation} I(z)=I_0\frac{L+z_0-z}{L+2z_0},  \label{8}
\end{equation}
here $z_0$ is a phenomenological parameter which, in the case of scalar waves for point-like isotropic scatterers, is determined by the photon mean free path $l$ as follows $z_0=0.7104l$. In our case, where the vector properties of the light are considered, Eq. (\ref{8}) is qualitatively confirmed. In reality, if we approximate the dependencies of Fig. 1 beyond the boundary region by a straight line then the obtained angles of inclination depend on detuning in accordance with the photon mean free path (compare with \cite{FKSH11}). Maximal angle of inclination corresponds to minimal $l$.

In addition to the frequency dependence, Eq. (\ref{8}) predicts the dependence of light intensity on slab thickness. In Fig. 2 we show the spatial distribution of atomic excitation in the clouds with different lengths. The atomic density in both clouds is the same and equals to $n=0.3$.  The radii of the two cylindrical clouds are also the same, $R=25$; external radiation is in resonance with the free atom transition, $\Delta=0$.
\begin{figure}[th]
\begin{center}
{$\scalebox{0.85}{\includegraphics*{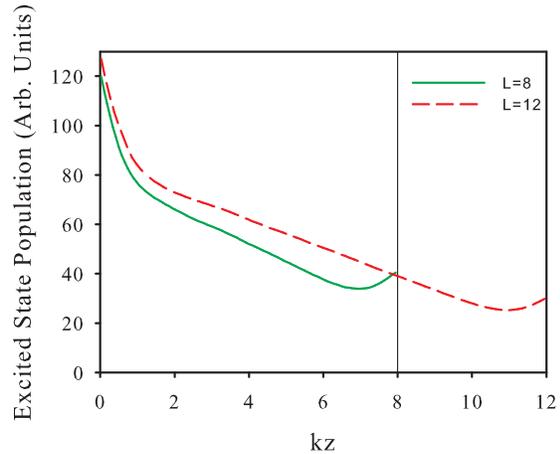}}$ }
\caption{Spatial distribution of atomic excitation in clouds of different length $L$. Atomic density is $n=0.3$.
Calculations were made for the central part $r<8$ of a cylindrical cloud with radius $R=25$.}
\end{center}
\par
\label{fig2}
\end{figure}

An increase in the length of the cloud leads to a decrease in the rate of total intensity attenuation. This decrease is in good qualitative agreement with Eq. (\ref{8}), especially taking into account the approximate nature of this formula and that in our case the length of the atomic ensemble exceeds the mean free path only several times.

One more peculiarity, which is clearly visible in Figs. 1 and 2, are the boundary effects.  Atoms located in the border region with a size of about $1 - 1.5$ on the scale of $kz$ interact predominantly with the atoms, which are situated to one side, inside the cloud. This causes some modification of the dipole-dipole interatomic interaction. Besides this, electromagnetic waves reflect from the interface between the vacuum and the cloud, which also influences the observed boundary effects.

These effects manifest themselves differently for different frequencies of external radiation. For some frequencies the population increases with the approach to the boundary, for others it diminishes.  For some frequencies, for example, where $\Delta=2 \gamma$, the dependence is nonmonotonic in the border region.

For the given frequency, the nature of the wall effects varies on the near and far edge with respect to the light source. On the nearer boundary, the polarization of the incident light plays an important role.  The latter is well demonstrated  by Fig. 3, in which the spatial distribution of the atoms, excited to the different Zeeman sublevels is shown for two frequencies of the external radiation.
\begin{figure}[th]
\begin{center}
{$\scalebox{0.85}{\includegraphics*{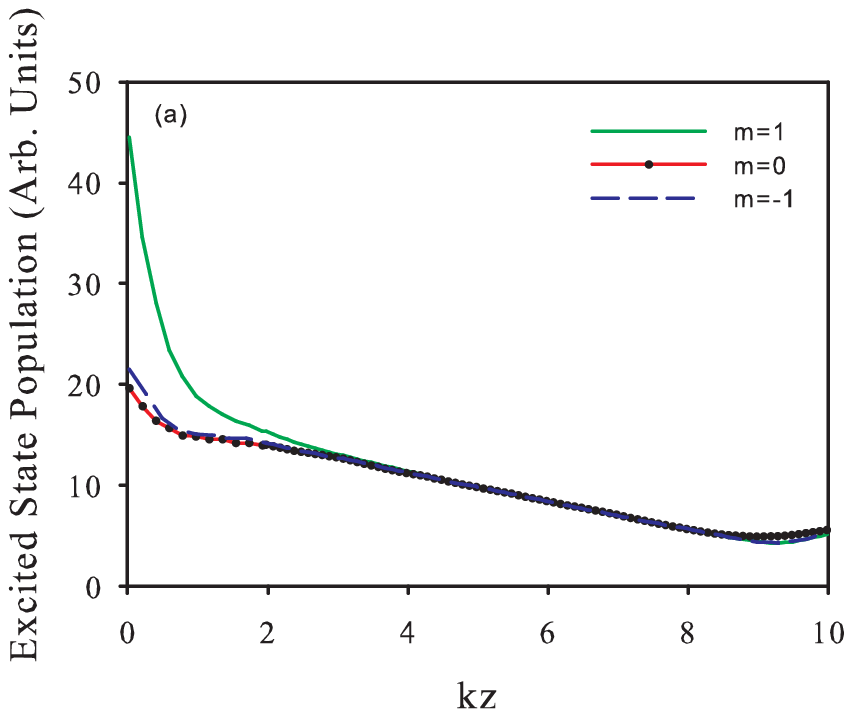}}$ }
{$\scalebox{0.91}{\includegraphics*{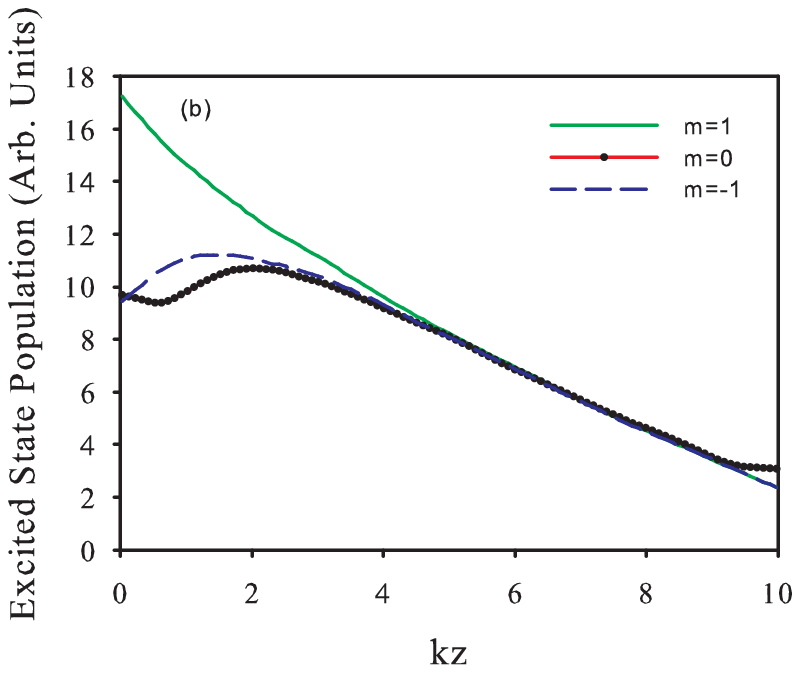}}$ }
\caption{Spatial distribution of different Zeeman sublevels population. $R=20$, $L=10$, $n=0.2$. (a) $\Delta=0$ and  (b) $\Delta=1.5\gamma$.  }
\end{center}
\par
\label{fig3}
\end{figure}

In the central parts of the cloud, far from the edges, the influence of nonscattered external radiation is small and the trapped radiation is unpolarized. Besides that, in the present case the atomic medium is isotropic on average. That is why averaged populations of different sublevel coincide with each other with a great accuracy. Equalization of populations occurs more rapidly as the photon mean free path decreases.  So for $\Delta=0$ the difference in population becomes negligible at $z\sim 2.5$, while for $\Delta=1.5\gamma$ it take place only for $z\sim 4$.

Near the far edge of the cloud the populations of sublevels $m=1$ and $m=-1$ are still the same, but they differ from that of sublevel  $m=0$. This difference can be explained through two factors. First, near the boundary the atomic ensemble ceases to be isotropic on average. A preferential direction appears, which is the direction of the perpendicular to the surface, which in our case coincides with the z direction. Orthogonal to the z axis direction the atoms are located symmetrically. The second reason which can cause unequal population over Zeeman sublevels is asymmetry of optical excitation. Trapped radiation illuminates atoms from the internal parts of the cloud. Besides that the reflection of the light from the interface between vacuum and cloud is accompanied by light polarization effects. The population of sublevel $m=0$ is determined by light linearly polarized along the z axis while sublevels $m=\pm 1$ are populated by the light which has polarization vector with a nonzero projection on the plane perpendicular to this axis.

Thus, in the edge regions, optical pumping effect takes place. In the region $z\sim L$, where the average value of atomic angular momentum is zero and the average value of squared angular momentum does not equal to zero the atomic ensemble is aligned on its angular momentum. At the front edge, along with alignment effects, orientation is also generated.

Fig. 1 shows the spatial distribution of atomic excitation for different external light detuning and for given atomic density. The dependence of this distribution on density for resonant light with $\Delta=0.5\gamma$ is given in Fig. 4.  The calculations are made for the region  $r<6$ of the clouds with radius  $R=15$ and length $L=10$.

\begin{figure}[th]
\begin{center}
{$\scalebox{0.85}{\includegraphics*{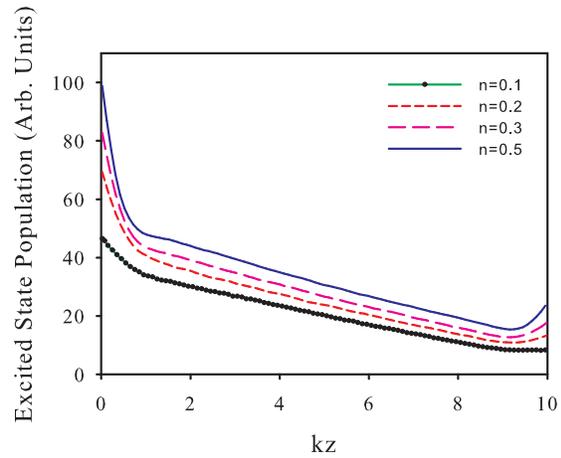}}$ }
\caption{Spatial distribution of atomic excitation in the clouds with different densities $n$. Probe radiation detuning is $\Delta=0.5\gamma$}
\end{center}
\par
\label{fig5}
\end{figure}

The essential feature of the results shown in Fig. 4 is the very weak dependence of the magnitude of atomic excitation on atomic density. A density increasing from $n = 0.1$ to $n = 0.5$ leads approximately to only a 60\% increase of the total number of excited atoms. This effect in our interpretation connects with level shifts caused by strong dipole-dipole interaction for dense media. These shifts are approximately proportional to the cube of the average interatomic separation for dense clouds (see (\ref{5})), i.e. they are proportional to atomic density. So inhomogeneous broadening of the absorption line is in direct proportion to density. This broadening compensates with increasing the total number of atoms, i.e. for a given size of cloud increasing in density.

In addition to these factors, increasing density does not change the nature of the spatial distribution of atomic excitation. Outside the boundary regions there is no significant deviation from linear dependence for any of the considered conditions even in spectral regions where, according to \cite{FKSH11}, the Ioffe-Regel criterium is satisfied.

To determine the deviation of spatial dependencies from linear we approximated the curves in Fig. 4 by a polynomial function of the third order. The approximation was performed for the central part of the cylinder from $z=2.0$ to $z=8.0$. As a result for $n=0.5$ and $\Delta=0.5\gamma$, we obtained $\rho_{exc}=53.8-4.98z+0.056z^2-0.0038z^3$.
The coefficient of $z^2$ is 90 times smaller than that of $z$. This allows us to make inference about the absence of manifestation of strong localization effects in the clouds with the considered parameters.

All spatial dependencies presented above are obtained as average values over random atomic configurations. Taking into account different spatial configurations of the atomic ensemble we model partially the influence of residual atomic motion which take place even for cold clouds in specialized traps. At the same time it is known \cite{MD} that under conditions close to localization the averaged value of such random quantities as transmission of atomic ensemble may differ essentially from its most probable value.  This difference is caused by the influence of rare configurations of scatterers which give a large contribution to the mean value. To consider the role of fluctuations in the problem under study we have analyzed the distribution law for atomic excitation.

In Fig. 5 we show the probability density of total population in the excited state of atoms in a unit volume. Magnitudes $\rho_{exc}$ were obtained as a sum over all Zeeman sublevels of the excited states. Averaging over $\Delta V=\pi R_0^2 \Delta z$ is performed in the limit of a cylindrical volume with radius $R_0=8$ and length $\Delta z=0.4$. Increasing the averaging volume with respect to previous calculations (Figs. 1-4 were obtained for $\Delta z=0.1$) was made to decrease the role of fluctuations connected with dispersion of the total number of atoms in the volume $\Delta V$. Because the dense clouds where Ioffe-Regel criterium is satisfied are most interesting for us we consider ensembles with $n=0.5$. For such density the Ioffe-Regel criterium is satisfied for a wide spectral region from $\Delta\sim 0$ to $\Delta\sim 4\gamma$ (see \cite{FKSH11}). The spectral domain  $0.8\gamma<\Delta< 1.8\gamma$ is the most interesting and promising with regards to unusual optical properties of atomic ensemble. Here not only is the Ioffe-Regel criterium satisfied but also the real part of dielectric constant is negative, i.e. polariton states are possible.  Curves of Fig. 5 are calculated for external radiation with detuning $\Delta=\gamma$.

\begin{figure}[th]
\begin{center}
{$\scalebox{0.85}{\includegraphics*{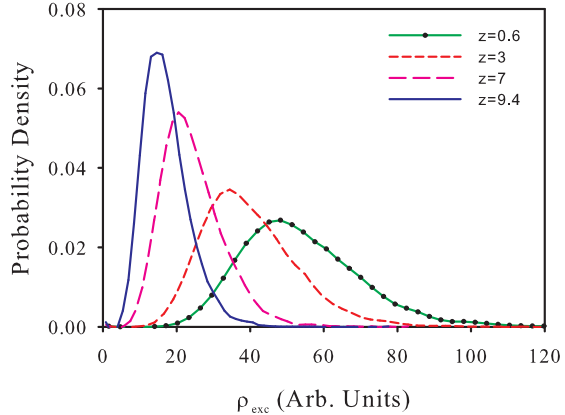}}$ }
\caption{The distribution law for atomic excitation in different parts of a cylindrical cloud with radius $R=20$. Calculation is performed for the central region  $r<8$ near the axis of the homogeneous (on average) atomic cloud.  Atomic cloud density is $n=0.5$, exciting radiation detuning is $\Delta=\gamma$. }
\end{center}
\par
\label{fig5}
\end{figure}

The distribution law for atomic excitation in Fig. 5 is calculated for different parts of the cylindrical cloud. All curves  are normalized to unit magnitude. From Fig. 5 it is clearly seen that the distribution functions of the random variable $\rho_{exc}$ are asymmetrical and have strongly pronounced tails. This asymmetry leads to difference between average and mean values of random variable. However this difference is not large. In Fig. 6 we show two pairs of curves depicting the spatial distribution of atomic excitation for two different detunings of external radiation. The first pair is calculated on the basis of the average value as done previously, and the second one is build on the most probable magnitudes. The quantitative difference in curves is clearly visible but qualitatively the behavior of these curves are identical. Small deviations from linear dependence lies within accuracy of the calculations and does not allow making conclusion about localization phenomenon.

\begin{figure}[th]
\begin{center}
{$\scalebox{0.85}{\includegraphics*{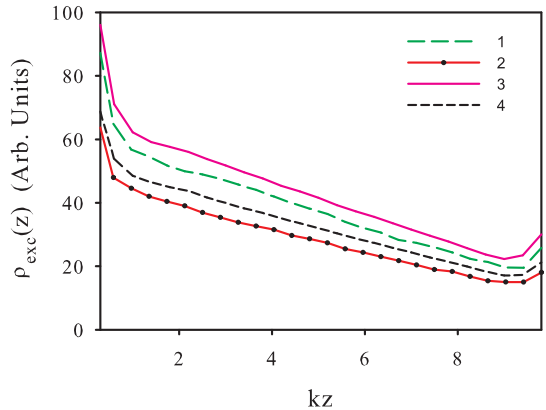}}$ }
\caption{Spatial distribution of excited atoms. Curves $1$ and $3$ are calculated for $\Delta=0$, curves $2$ and $4$ for $\Delta=\gamma$. Curves $3$ and $4$ correspond to average values of the excited state populations while $1$ and $2$ to the most probable values. The others parameters are as in Fig. 5.}
\end{center}
\par
\label{fig6}
\end{figure}

\subsection{Comparison of results from the vector and scalar dipole-dipole interactions}

The great majority of investigations of optical properties of cold atomic ensembles, particularly investigations of localization phenomena, were performed in the framework of the so called scalar theory. To show that predictions of such theory may contradict essentially to the calculation where vector nature of electromagnetic field are taken into account we calculated dependencies shown in Fig. 6 but in the scalar approximation. Corresponding results are given in Fig. 7.

\begin{figure}[th]
\begin{center}
{$\scalebox{0.85}{\includegraphics*{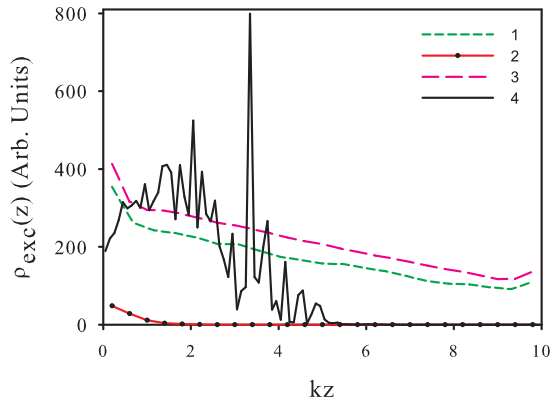}}$ }
\caption{The same as in Fig. 6 but all calculations are made in the framework of the scalar approximation.}
\end{center}
\par
\label{fig7}
\end{figure}

As in Fig. 6 here we show two pairs of curves calculated both on the basis of average values and on most probable ones. For resonant radiation with $\Delta=0$ any effects of localization do not practically appear \cite{SS}. But for $\Delta=\gamma$ the situation changes dramatically. First of all, the density of excited atoms decreases very rapidly inside the cloud. For $z>5$ we have practically no excited atoms. Second, huge fluctuations takes place here. These fluctuations manifest themselves even after averaging over a large number of statistical tests. The curve labelled as $4$ in Fig. 7 is obtained after averaging over 10,000 random spatial configurations of atoms in the ensemble. The root-mean-square deviation here exceeds the mean value several dozen times. While for $\Delta=0$ this deviation is much smaller and determined primarily by fluctuation of the total number of atoms in the averaging volume.

Another peculiarity of the atomic excitation by radiation with detuning $\Delta=\gamma$ predicted by the scalar model is the essential difference between the averaged excitation and the most probable one (curves $2$ and $4$ in Fig. 7). Here we deal with a feature of the transmission of dense cloud in the regime of strong localization mentioned in \cite{MD}). This feature is confirmed by the results of Fig. 8.

\begin{figure}[th]
\begin{center}
{$\scalebox{0.85}{\includegraphics*{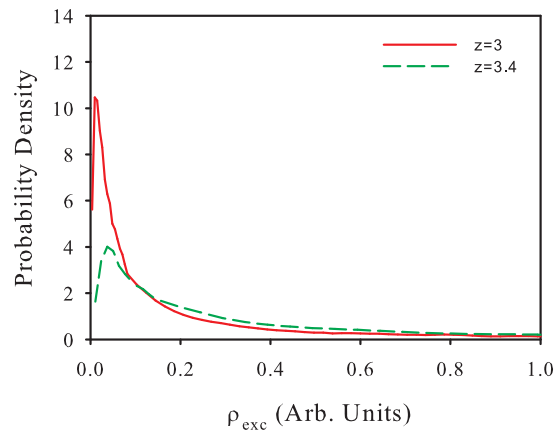}}$ }
\caption{The distribution law for atomic excitation in different parts of a cylindrical cloud. Scalar approximation.}
\end{center}
\par
\label{fig8}
\end{figure}

In Fig. 8, we show the distribution law for atomic excitation calculated in the scalar approximation. As the z coordinate of the point of observation increases the distribution becomes more and more narrow. The most probable value moves toward very small quantities. At the same time, a very long tail appears. This tail corresponds to rare and strong atomic excitation connected with random large values of transmission.

\section{Conclusions}
In the present paper we analyze the influence of the resonant dipole-dipole interatomic interaction in dense atomic clouds on the spatial distribution of atoms excited by weak monochromatic external radiation. The population of different Zeeman sublevels of the excited state are calculated. The dependence of this population distribution on atomic density as well as on the frequency of external radiation is studied. In all considered cases, beyond the spatial boundary regions, the total density of excited atoms decreases linearly inside the atomic ensemble. Near the boundary regions, excited state orientation and alignment of the atomic angular momentum is obtained; the various effects are due to a combination of coherent optical excitation, diffusive excitation, and the spatial asymmetry of the optical excitation near the sample boundaries. Manifestations of strong localization effect were not found even in those parameter regions where Ioffe-Regel criterium of strong localization is satisfied. This result is consistent with recent measurements \cite{Mark_exp,Mark_exp2} in high density $^{87}Rb$, which showed no clear evidence for light localization in spite of the density being sufficiently high that the Ioffe-Regel criterion was satisfied.  However, comparative calculations performed in the framework of the scalar approximation display explicit manifestation of strong localization for some conditions.  This result is in good agreement with deductions made in a recent paper \cite{SS} on the analysis of the eigenvalues of the matrix $\Sigma _{e_a^m e_b^{m'}}$ (\ref{5}) and (\ref{6}).

\subsection*{Acknowledgements}
We thank Professor D. V. Kupriyanov for fruitful discussions. We appreciate the financial support of the Federal Program for Scientific and Scientific-Pedagogical Personnel of Innovative Russia for 2009-2013 (Contract No. 14.B37.21.1938).  We also acknowledge financial support of the National Science Foundation (Grant Nos. NSF-PHY-0654226 and NSF-PHY-1068159).


\baselineskip18 pt

\begin{thebibliography}{99}
\bibitem{Cao} Hui Cao, \emph{Lasing in Disordered Media}, in Progress in Optics 45,
(2003).

\bibitem{Wiersma} D.S. Wiersma, Nature Phys. 4, 359 (2008).

\bibitem{Conti} C. Conti and A. Fratalocchi, Nature Phys. 4, 794 (2008).

\bibitem{KaiserRandom1} L. Froufe-P\`{e}rez, W. Guerin, R. Carminati and R. Kaiser Phys. Rev.
Lett.102, 173903 (2009).

\bibitem{KaiserRandom2} W. Guerin, N. Mercadier, D. Brivio and R. Kaiser,
Optics Exp. 17, 14 (2009).

\bibitem{3} D. Bouwmeester, A. Ekert, A. Zeilinger, \emph{The Physics
of Quantum Information}, Springer-Verlag, Berlin, Germany, (2001).

\bibitem{4} M.D. Lukin, Rev. Mod. Phys. 75, 457 (2003).

\bibitem{5} P.W. Milonni, \emph{Fast Light, Slow Light, and Left-handed Light},
Taylor and Francis, New York, (2005).

\bibitem{6} M. Fleishhauer, A. Imamoglu, and J.P. Marangos, Rev. Mod. Phys.
77, 633 (2005).

\bibitem{7} L. V. Hau, Nature Photonics 2, 451 (2008).

\bibitem{8} D.A. Braje, V. Balic, G.Y. Yin, S.E. Harris, Phys. Rev. A 68,
041801 (2003).

\bibitem{9} S. Ospelkaus, A. Peer, K.-K. Ni, J. J. Zirbel, B. Neyenhuis, S.
Kotochigova, P. S. Julienne, J. Ye, and D. S. Jin, Nature Phys. 4, 622
(2008).

\bibitem{10} G. K. Campbell, A. D. Ludlow, S. Blatt, J. W. Thomsen,M. J.
Martin, M. H. de Miranda, T. Zelevinsky, M. M. Boyd, J. Ye, S. A. Diddams, T.
P. Heavner, T. E. Parkerand S. R. Jefferts, Metrologia 45, 539 (2008).

\bibitem{11} Jun Ye, S. Blatt, M. M. Boyd, S. M. Foreman, E. R. Hudson,
Tetsuya Ido, B. Lev, A. D. Ludlow, B. C. Sawyer, B. Stuhl, T. Zelinsky, Int.
J. Modern Phys. D 16, 2481 (2007).

\bibitem{24} D.S. Wiersma, P. Bartolini, Ad Lagendijk, R. Righini,
Nature 390, 671 (1997).

\bibitem{25} A.A. Chabanov, M. Stoytchev, A.Z. Genack, Nature 404, 850
(2000).

\bibitem{26} M. Storzer, P. Gross, C.M. Aegerter, G. Maret, Phys. Rev.
Lett. 96, 063904 (2006).

\bibitem{27} C.M. Aegerter and G. Maret, \emph{Coherent backscattering and
Anderson localization of light}, in Progress in Optics 52, 1 (2009).

\bibitem{Mark_exp} S. Balik, A.L. Win, M.D. Havey, I.M. Sokolov and D.V. Kupriyanov,
 to appear, Physical Review A (2013); arXiv:0909.1133 (2009).

\bibitem{Mark_exp2} S. Balik, A.L. Win, M.D. Havey, A.S. Sheremet, I.M. Sokolov, and D.V. Kupriyanov,
arXiv:1303.0037 (2013).

\bibitem{Kaiser} R. Kaiser, J. Mod. Optics 56, 2082 (2009).

\bibitem{FKSH11} Ya. A. Fofanov, A.S. Kuraptsev, I.M. Sokolov, and M.D. Havey,
Phys. Rev. A, 84, 053811 (2011).

\bibitem{SKH11} I.M. Sokolov, D.V.  Kupriyanov and M.D. Havey,
\textit{J. Exp. Theor. Phys.} \textbf{112}, 246 (2011).

\bibitem{SKKH09} I.M. Sokolov, M.D. Kupriyanova, D.V. Kupriyanov, and M.D.
Havey, Phys. Rev. A 79, 053405 (2009).


\bibitem{Ross} M. C. W. van Rossum and Th. M. Nieuwenhuizen, Rev. Mod. Phys. 71, 313 (1999).

\bibitem{MD} Cord A. Muller and Dominique Delande, $\emph{Disorder and interference: localization
Phenomena}$ arXiv:1005.0915.

\bibitem{SS} Sergey E. Skipetrov and Igor M. Sokolov, arXiv:1303.4655 (2013).

\end{thebibliography}
\end{document}